\documentstyle[12pt]{article}
\begin{document}
\renewcommand{\thesection}{\Roman{section}}
\newcommand{\rt}{\rightarrow}
\newcommand{\en}{\begin{eqnarray}}
\newcommand{\ed}{\end{eqnarray}}
\baselineskip 17pt
{\hfill hep-ph/xxyyzzz}
\vskip 3cm
\begin{center}
{\large\bf Quarkonium production in SM Higgs decays}
\end{center}
\vskip 7mm
\centerline{Cong-Feng Qiao}
\centerline{\small\it China Center of Advanced Science and Technology
(World Laboratory),}
\centerline{\small \it Beijing 100080, People's Republic of China}
\centerline{\small \it and I.C.T.P., P.O. Box 586, 34100 Trieste, Italy}
\vskip 3pt
\centerline{Feng Yuan,~and Kuang-Ta Chao}
\centerline{\small\it Department of Physics, Peking University, 
            Beijing 100871, People's Republic of China}
\centerline{\small\it and China Center of Advanced Science and Technology
(World Laboratory),}
\centerline{\small \it Beijing 100080, People's Republic of China}
\begin{center}
\begin{minipage}{130mm}
\vskip0.6in
\begin{center}{\bf Abstract}\end{center}
  { We investigate the SM Higgs decays into heavy quarkonia $J/\psi$ 
and $\Upsilon$ in both color-singlet and color-octet mechanisms. It is
found that in $J/\psi$ production the contribution through color-octet
processes overwhelm that in color-singlet processes all the intermediate
region $M_Z<M_{H^0}<2 M_w$ and the fraction ratio is comparatively large,
but in $\Upsilon$ production the contribution through color-octet
mechanism is negligible.}
\vskip 1cm
PACS number(s):12.38.Bx, ~13.38.Dg, ~14.40.Gx
\end{minipage}
\end{center}

\vfill\eject\pagestyle{plain}\setcounter{page}{1}
Despite the success of the Standard Model (SM) \cite{p0} in describing
elementary particle phenomena at electroweak energy scale and below, the
theory has some intrinsic shortcomings which may not be overcome in the
context of the theory itself. For example, the hierarchy problem existed
in the gauge symmetry breaking. The mechanism of spontaneous symmetry
breaking introduced in SM to allow some of the gauge bosons and fermions
to acquire mass while keeping the theory renormalizable leads to a neutral
scaler particle, the Higgs boson $H^0$. While a worldwide experimental
effort has been made, this Higgs particle has not yet been discovered.

Though the SM predicts the existence of the  $H^0$ with unspecified mass,
the couplings of it with other fundamental particles are well defined.
Consequently the $H^0$ production cross section in various channels and
the decay branching ratios of it are precisely predicted as a function of
the Higgs boson mass. According to the LEP I data, based on the Bjorken
process $e^+ e^- \rightarrow H^0 Z^*$, the Higgs mass, $M_{H^0}$, should
be heavier than $77.1$ GeV \cite{p1}.

The hunt of the Higgs boson meets two problems, i.e., how to produce it and
then how to detect it. Therefore, experimental researches at LEP include
both $H^0$ production and  decays, meanwhile the search range will expand
with the operation of LEP at higher energy. After the operation of the
Large Hadron Collider (LHC), almost all the possible energy
region where Higgs may exist will be searched down.
In this paper, we proceed out discussion in the intermediate energy 
range $M_Z<M_{H^0}<2 M_w$, which is the most difficult range in experiment
to detect the Higgs boson.

In theory, the study for the nature of the Higgs boson have been in 
highlight in particle physics ever since the establishment of SM.
The probabilities of finding the $H^0$ and other Higgs bosons
appearing in theories beyond the minimal SM in various
collider facilities are evaluated \cite{p2}\cite{p3}. Among others of 
Higgs boson production processes, the Yukawa process, i.e., the
bremsstrahlung production of the neutral Higgs boson $H^0$ from a heavy
fermion, of the SM Higgs boson is an important channel to discover
and to investigate this particle. In experiment, the analysises of the
Yukawa process have led to some constraints on the theories in respect to
the Higgs sector \cite{3pp}. Among various decay processes of the Higgs boson,
Fermionic decay channels of it are important channels.
The observation of such decay modes may provide important information on
fermion Yukawa couplings which associate with the symmetry breaking
mechanism in the SM. The subprocess of the fermionic decay channels, 
that is quarkonium production in the Higgs boson decays that may be called
the Extended Wilczek Processes (EWP), which helps to further identify 
the Higgs boson signals also gains much attention though
there may be large theoretical uncertainties both due to the 
Quantum Chromodynamics (QCD) and relativistic corrections \cite{p2}
\cite{p4}. Considering the fact that the low limit of the SM Higgs mass
is up to  about the mass of $Z^0$ boson's and the recent developments in
quarkonium production theory, in this paper we reevaluate the EWP within
the scope of Standard Model. 

In the past few years, two approaches in describing the heavy quarkonium
production, the fragmentation mechanism \cite{p5} and the Non-Relativistic
Quantum Chromodynamics (NRQCD) factorization \cite{p6}, have greatly
improved the theoretical studies in this area. The introduction of
color-octet mechanism asks ones to reevaluate almost all the processes
associated with quarkonium production, in order either to test
the new production mechanism, to further understand QCD or to assess the
quarkonium production rates in a specific process. Of course, the EWP is
also in the case.

For charmonium and bottomonium production in the Higgs boson decay,
the important color-singlet channels include $H^0\rightarrow V \gamma$
(electromagnetism), $H^0\rt Q\bar{Q} V $ (quark fragmentation),
$H^0\rt b \bar{b} V$ (photon fragmentation), and $H^0\rt b\bar{b} V gg$
(gluon fragmentation) processes as shown in Fig.1, and the process $
H^0\rt V gg$ vanishes. Here, $V$ represents  a vector meson consisting of
a $Q\bar{Q}$ pair ($Q$ is  heavy quark $b$ or $c$). 

The decay width of the process of Fig.1(a) can be calculated 
straightforward making use of the standard formalism \cite{p8}.
\begin{eqnarray}
\label{e1}
\Gamma =\frac{\alpha e_Q^2}{27 v^2} \xi_V (1-\xi_V^2)
    <0|{\cal O}_1^{V} (^3S_1)|0>,
\end{eqnarray}
where $e_Q$ is the charge fraction of heavy quark, $e_c=\frac{2}{3}$
and $e_b=-\frac{1}{3}$; $v$ is the characteristic quantity which
sets the scale of spontaneous symmetry breaking in $SU(2)_L \times
U(1)$ theory and $v\simeq 246$ GeV; $\xi_V$ is defined as $\xi_V=
\frac{M_V}{M_{H^0}}$, $M_V$ and $M_{H^0}$ are the masses of the bound systems
and the SM Higgs boson; $<0|{\cal O}_1^{V} (^3S_1)|0>$ is the matrix
element of color-singlet four fermion operator of quarkonium, which is
associated with the nonperturbative part of the quarkonium production and
its value may be determined by quarkonium electronic decay widths or from
the potential model. Because the Higgs $H^0$ couples preferentially to the
heavy fermion, when $M_{H^0}< 2 M_w $, $M_w$ is the mass of $W$ boson,
the dominant decay channel in Higgs decays is $H^0\rt b\bar{b}$ even
considering the virtual $Z^*$ and $W^*$ process \cite{xxx}.
So, here after, we will constrain the $H^0$ coupling only to the 
quarks ($b,~c$) in our discussion.

\en
\label{e2}
\Gamma_{H^0\rt b\bar{b}} = 
\frac{3m_b^2}{8\pi v^2} M_{H^0} (1-4 \eta^2)^{3/2},
\ed
where $\eta\equiv m_b/M_{H^0}$.
From Eq.(\ref{e1}) and Eq.(\ref{e2}), the fraction ratio
\en
\label{e3}
R_{J/\psi}\equiv \frac{\Gamma_{H^0\rt J/\psi\gamma}}{\Gamma_{H^0\rt
b\bar{b}}}
=\frac{32 \pi \alpha}{729 M_{H^0}
m_b^2}\frac{\xi_{J/\psi}(1-\xi_{J/\psi}^2)}{(1-4\eta^2)^{3/2}}
 <0|{\cal O}_1^{J/\psi} (^3S_1)|0>,
\ed
\vskip -0.5mm
\en
\label{e3p}
R_{\Upsilon}\equiv\frac{\Gamma_{H^0\rt \Upsilon\gamma}}{\Gamma_{H^0\rt
b\bar{b}}}
=\frac{8\pi \alpha}{729 M_{H^0}
m_b^2}\frac{\xi_\Upsilon (1-\xi_\Upsilon^2)}{(1-4\eta^2)^{3/2}}
 <0|{\cal O}_1^{\Upsilon} (^3S_1)|0>\nonumber \\\approx 
\frac{16 \pi \alpha}{729 M_{H^0}^2 m_b}(1-\xi_\Upsilon^2)^{-1/2}
 <0|{\cal O}_1^{\Upsilon} (^3S_1)|0>, 
\ed
\vskip 0.5mm
\par
\noindent
can then be easily obtained. The decay $H^0\rt V \gamma$ with the subsequent
decay $V\rt e^+ e^-$ from experimental point of view presents the clearest
signal in searching the Higgs boson. Unfortunately, the ratios of these decay
rates to the decay rate $H^0\rt b\bar b$ is too small (see below and
Ref.\cite{yyy}).

Because the minimal value of the mass of $H^0$ is over $77.1$ GeV, the
decay widths, therefore the fraction ratios of the process shown in
fig.1(b) can be evaluated using the universal quark fragmentation
functions \cite{p9}.
\en
\label{e4}
R_{J/\psi}\equiv \frac{\Gamma(H^0\rightarrow c\bar cJ/\psi)}{\Gamma(H^0
\rightarrow b\bar b)}
=2\times P_{c\rt J/\psi}\times \frac{m_c^2}{m_b^2},
\ed
where $P_{c\rt J/\psi}$ is the universal fragmentation probability, and
\begin{eqnarray}
\label{e5}
P_{c\rt J/\psi} = \frac{16}{243}
\alpha_s(2m_c)^2 \frac{
 <0|{\cal O}_1^{J/\psi} (^3S_1)|0>
}{m_c^3}(\frac{1189}{30}-57 ln2).
\end{eqnarray}
As Eq.(\ref{e4}),(\ref{e5}), 
the fraction ratio of $\Gamma_{H^0\rt b \bar{b} \Upsilon}$ over
$\Gamma_{H^0\rt b \bar{b}}$ can be 
readily obtained,
\begin{eqnarray}
\label{e6}
R_{\Upsilon}\equiv \frac{\Gamma(H^0\rightarrow b\bar b\Upsilon
)}{\Gamma(H^0
\rightarrow b\bar b)}
=2\times P_{b\rt \Upsilon}.
\end{eqnarray}
Here $P_{b\rt\Upsilon}$, the fragmentation probability of $b\rt \Upsilon$,
is in the same form as Eq.(\ref{e5}),
\en
\label{e7}
P_{b\rt \Upsilon } = \frac{16}{243}
\alpha_s(2m_b)^2 \frac{
 <0|{\cal O}_1^{\Upsilon} (^3S_1)|0> 
}{m_b^3}(\frac{1189}{30}-57 ln2).
\ed

The process as shown in Fig.1(c) can be calculated using the standard
method in computing the quarkonium production and decays and the result
is simplified into the following form after neglecting the mass of the
$b(\bar b)$ quark (antiquark), except for in the Higgs-quark coupling
vertex. 
\en
\label{e9}
\frac{d \Gamma(H^0\rt \rm{V} b \bar{b})}{dz_1 dz_2}=\frac{\alpha^2 m_b^2
e_Q^4 M_{H^0}}{54 M_V^3 v^2\pi}\frac{<0|{\cal O}_1^{V} (^3S_1)|0>}{z_1^2
z_2^2} [(z_1 z_2-\xi_V^2)(z_1^2+ z_2^2)+ 2 z_1 z_2 (z_1-1)(z_2-1)],
\ed
where $z_1=1-\frac{2k\cdot p_2}{M_{H^0}^2}$, $z_2=1-\frac{2k\cdot p_1}
{M_{H^0}^2}$ and their integration regions are $r\leq z_1 \leq 1$,
$r/z_1\leq z_2\leq 1+r -z_1$. $k,~p,~p_1,~p_2$ are the momenta of the
Higgs boson, bound state, and the final quarks appearing in Fig.1(c).
Because Eq.(\ref{e9}) is a differential equation, it can be
used to evaluated the energy distribution of quarkonia, which is different
from that of the two-body decay mode. After integrating over $z_1$ and
$z_2$, the total decay width can be obtained and hence the fraction
ratios.

The decay width of the process $H^0\rt b\bar{b} V gg$ as shown in Fig.1(d)
can be expressed as 
\en
\label{e10}
\Gamma(H^0 \rt b \bar{b} g^*, g^*(\mu)\rt gg V )=
\int\limits^{M_{H^0}^2}_{M^2}d\mu^2 \Gamma (H^0\rt b \bar{b} g^*(\mu))
P_{g^* \rt gg V }.
\ed
Here, $P_{g^*\rt gg V}$ 
is the decay distribution of the virtual gluon with virtuality
$\mu$, and it is defined as 
\en
\label{e11}
\Gamma(g^*\rt ggV) = \pi \mu^3 
P_{g^* \rt gg V }.
\ed
The expression of $\Gamma(g^*\rt gg V)$ can be found in Ref.\cite{p10},
the $P_{g^* \rt gg V }$, therefore, may be induced.
\begin{eqnarray}
\label{tt1}
&&\mu^2 P(g^*\rightarrow V g g)=
C_V r \int\limits_{2\sqrt{r}}^{1+r} dx_{V}
 \int\limits_{x_{-}}^{x_{+}}dx_{1} f(x_{V},x_1;r),
\end{eqnarray}
where $r\equiv M_V^2/{\mu}^2$, 
and
\begin{eqnarray}
\label{uu}
C_{J/\psi}=\frac{10\alpha_s^3}{243\pi^2}\frac{
 <0|{\cal O}_1^{J/\psi} (^3S_1)|0> 
}{M_{J/\psi}^3},
~~C_{\Upsilon}=\frac{10\alpha_s^3}{243\pi^2}\frac{
 <0|{\cal O}_1^{\Upsilon} (^3S_1)|0> 
}{M_{\Upsilon}^3}.
\end{eqnarray}
The function $f$ in Eq.(\ref{tt1}) is of the form 
\begin{eqnarray}
\nonumber
f(x_{V},x_1;r)&=&\frac{(2+x_2)x_2}{(2-x_{J/\psi})^2(1-x_1-r)^2} +
 \frac{(2+x_1)x_1}{(2-x_{J/\psi})^2(1-x_2-r)^2}\\ 
 \nonumber
 &+&
 \frac{(x_{J/\psi}-r)^2-1}{(1-x_2-r)^2(1-x_1-r)^2}
+ \frac{1}{(2-x_{J/\psi})^2}\Big(\frac{6(1+r-x_{J/\psi})^2}
 {(1-x_2-r)^2(1-x_1-r)^2}\\ 
& + & \frac{2(1-x_{J/\psi})(1-r)}{(1-x_2-r)(1-x_1-r)r}
  +\frac{1}{r}\Big ),
\end{eqnarray}
where 
$x_{i}\equiv 2E_{i}/\mu$ with $i=V,~g_1,~g_2$ are the energy fractions
carried by the quarkonia and two gluons in the $g^*$ rest frame, and then 
$x_2 = 2 - x_1 - x_{J/\psi}$. The limits of the $x_1$ integration in
Eq.(\ref{tt1}) are
\begin{eqnarray}
x_{\pm}= \frac{1}{2}(2 - x_{V} \pm \sqrt{x_{V}^2-4 r}) .
\end{eqnarray}
$\Gamma(H^0\rt b \bar{b} g^*(\mu))$ is the same as Eq.(\ref{e9}) with just
$M_V$ being changed to $\mu$. Therefore, the fraction ratio 
$H^0\rt b \bar{b} V gg$ to $H^0\rt b \bar{b}$ then can be calculated using
Eqs.(\ref{e2}) and (\ref{e10}). 

The heavy quarkonium production color-octet mechanism provides the 
process $H^0\rt Q\bar{Q}(^3S_1,\b 8) g \rt V  g $ with soft hadrons,
as shown in Fig.2(a), existing as the leading order process in $\alpha_s$.
Although this process makes a less significant contribution to the
$J/\psi$ production in $Z^0$ decays \cite{p7}, comparing with other higher
order processes, it may play a relatively more important role in the case
of $H^0$ decays. On the other hand, after summing over the soft hadrons
that accompany the quarkonium $V$ and over the collinear hadrons that make
up the gluon jet, the decay rate can be factored into the product of a
two-body decay rate and a nonperturbative matrix element, and the two-body
decay mode has a unique distributions in the kinematic variables of final
states relative to that of three-body's. Another color-octet process which
might play an important role in charmonium production in $H^0$ decays is
shown in Fig.2(b).

Although the nonperturbative color-octet matrix elements in these color-octet
processes are suppressed by $v^4$ with respect to that of color-singlet
processes according to NRQCD velocity scaling rules, the strong coupling
constant $\alpha_s$ greatly enhanced them relative to the corresponding
electromagnetic processes. The calculation of the decay widths of these
color-octet channels is similar to the corresponding color-singlet
processes, with only the change of coupling constant, the color factor,
and the nonperturbative matrix elements. Therefore, $H^0$ decay widths of
the processes as shown in Fig.2 can be inferred from Eq.(\ref{e1}) and
Eq.({\ref{e9}). 
\begin{eqnarray}
\label{e13}
\Gamma(H^0\rt V g \rm{X}) =\frac{\alpha_s }{3 v^2} \xi_V (1-\xi_V^2)
    <0|{\cal O}_8^{V} (^3S_1)|0>.
\end{eqnarray}
Here, the X stands for the soft hadrons produced in the neutralization
process of the color-octet $(Q\bar{Q})_8$ evolving into the heavy
quarkonium. The differential decay width of the Fig.2(b) process reads
as \en
\label{e14}
\frac{d \Gamma(H^0\rt \rm{V} b \bar{b} X)}{dz_1 dz_2}=\frac{\alpha_s^2
m_b^2 M_{H^0}}{12 M_V^3 v^2\pi}\frac{<0|{\cal O}_8^{V} (^3S_1)|0>}{z_1^2
z_2^2} [(z_1 z_2-\xi_V^2)(z_1^2+ z_2^2)+ 2 z_1 z_2 (z_1-1)(z_2-1)],
\ed
where the integration variables $z_1$ and $z_2$ change in the same regions as
that in Eq.(\ref{e9}). In the above $<0|{\cal O}_8^{V} (^3S_1)|0>$ is the
matrix element of four fermion color-octet operator. Its value may be
determined by the fitting of theoretical predictions to the experimental
data or from the lattice QCD calculations. Note that in the process of
Fig.2(a) the color-octet $Q\bar{Q}(^1S^{(8)}_0)$ and  
$Q\bar{Q}(^3P_J^{(8)})$ have zero contributions to the $H^0$ decay widths. 
From Eqs.(\ref{e2}), (\ref{e13}), and (\ref{e14}) the fraction ratios
of quarkonium production over b-quark pair production then can be
obtained.

In the numerical calculation, we take \cite{p9} 
\en
\nonumber
\alpha_s(2m_c)=0.26,~~ \alpha_s(2 m_b) =0.19,
\ed
\vskip -1.15cm
\en
\nonumber
M_{J/\psi}=2 m_c,~~M_{\Upsilon}= 2 m_b,~~ m_b=4.9~GeV,~~ m_c= 1.5~GeV,
\ed
and \cite{p14} 
\en
\nonumber
 <0|{\cal O}_1^{J/\psi} (^3S_1)|0>=1.2~GeV^3,~~
 <0|{\cal O}_1^{\Upsilon} (^3S_1)|0>=9.3~GeV^3,
\ed
\vskip -1.15cm
\en
\label{e15}
 <0|{\cal O}_8^{J/\psi} (^3S_1)|0>=6.6\times 10^{-3}~GeV^3,~~
 <0|{\cal O}_8^{\Upsilon} (^3S_1)|0>=5.9\times 10^{-3}~GeV^3.
\ed
With the mass of SM Higgs boson changing from 65 GeV to 155 GeV, the
relative weights and the magnitudes of the total fraction ratios of
$R_{J/\psi}$ and $R_{\Upsilon}$ are shown in Fig.3 and Fig.4. There are 
multiple components to each of the color-octet and -singlet
contributions in these figures. For the convenience in comparing the
relative importance of those diagrams in Fig.1 and Fig.2, we plot each
component of the octet and singlet in Fig.3 and Fig.4. Each figure of
them in fact contains nine curves corresponding to six subprocesses of
Fig.1 and Fig.2, the sum of both octet and singlet subprocess, and the
total. However, because the diversities of each component are so large
that some curves overlap, which means the dominant contribution is
almost the same as total.   

From the results shown in Fig.3 and Fig.4, it is interesting to see 
that the heavy quarkonium production in MS Higgs decays has two 
distinct features. First, generally, within the color-singlet model the
$\Upsilon$ production rate is smaller than that of $J/\psi$'s at collider
facilities, but in the SM Higgs decays where they have a comparable
fraction ratios because of the fact that the Higgs tends to couple with
heavy fermions. From Fig.3 and Fig.4 we can see that the process Fig.1(b)
always takes the leading contribution among all color-singlet ones
no matter in $J/\psi$ production or $\Upsilon$ production. 

Second, the color-octet mechanism makes a dominant contribution in the
$J/\psi$ production, however it makes a negligible contribution in the
$\Upsilon$ production. This is because that for the intermediate mass
Higgs the color-octet decay mode in Fig.2(b) always overwhelms that in
Fig.2(a), but in Fig.2(b) gluon propagator in $\Upsilon$ production 
suppresses the decay width relative to the $J/\psi$ production. This unique
character provides the EWP of $J/\psi$ production more reachable in
experiment than $\Upsilon$ production if the color-octet mechanism does
really work. In other words, if we would have measured a larger
production rate in $\Upsilon$ production than in $J/\psi$ in the Higgs
decays, we may conclude that the color-singlet mechanism is the dominant
one in quarkonium production, otherwise, the other production mechanism,
e.g., the color-octet one, must have taken part in. 

For the LHC luminosity of $\int {\cal{L}}=10^5 pb^{-1}$ there will yield
${\cal{O}}(0.2-1 \times 10^7)$ SM Higgs events in the mass range 155 GeV
$> M_{H^0} >$ 65 GeV with c.m. energy $\sqrt{s}=14$ TeV \cite{last}. From
this the quarkonium events produced at the proton collider LHC can be
readily estimated. For instance, at a mass of $M_{H^0}=100$ GeV, from
Fig.3 and Fig.4 we can read that the $R_{J/\psi}$ and $R_\Upsilon$ to be
about $10^{-4}$ and $2\times 10^{-5}$, from Ref.\cite{last} we get to know
that the  $H\rt b\bar{b}$ branching ratio is about 80\%. So, at this point
the cross section times the branching ratio for $p{p} \rt H(\rt J/\psi,
\Upsilon) + \rm{X}$ will yield about 200 $J/\psi$ and 80 $\Upsilon$
events.


In conclusion, after taking the color-octet mechanism into consideration
the quarkonium, especially the charmonium, decay mode of SM Higgs may
be detectable in the next generation collider, the LHC. That means 
the concerned quarkonium production process discussed in this paper may
stand as an implement to the two-photon decay mode in identifying the SM
Higgs in the intermediate mass region.

\vskip 1cm
\begin{center}
\bf\large\bf{ACKNOWLEDGEMENTS}
\end{center}

C.-F. Qiao thanks ICTP for the kind invitation for a visit, while this
work is completed. This work was supported in part by the Hua Run
Postdoctoral Science Foundation of China, National Natural Science
Foundation of China, the State Education Commission of China, and the
State Commission of Science and Technology of China.

\newpage
\centerline{\bf \large Figure Captions}
\vskip 1cm
\noindent
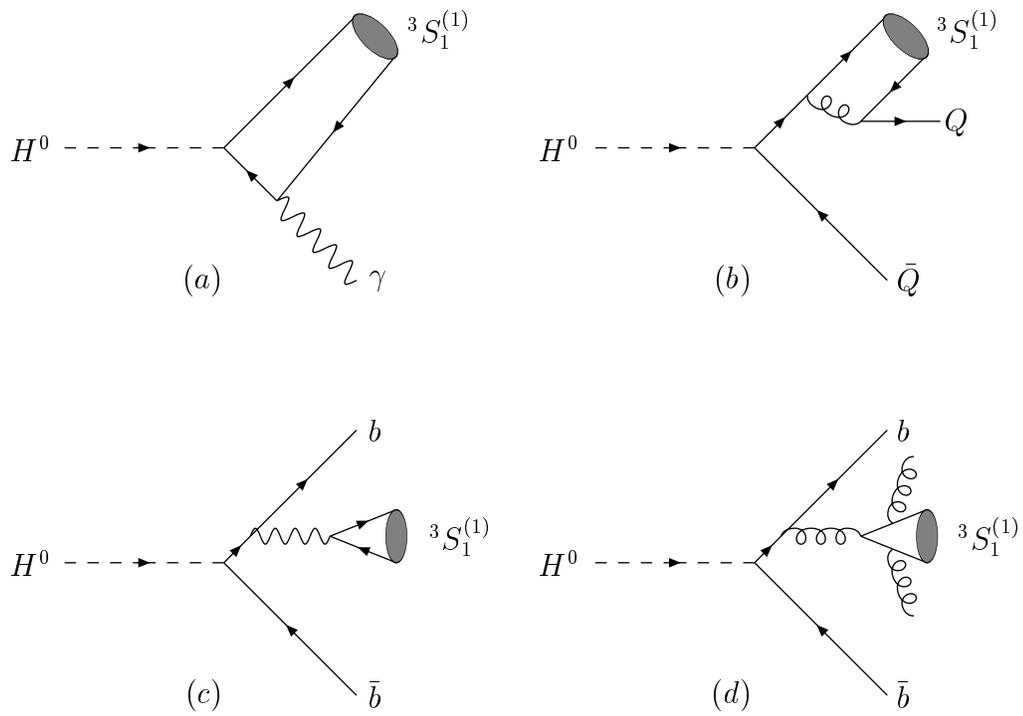
Fig.1. Color-singlet processes of $H^0$ decays into quarkonia, 
(a) electromagnetism process (b) quark fragmentation process (c) the
photon fragmentation process (d) gluon fragmentation process.\\
\vskip 0.2cm
\noindent 
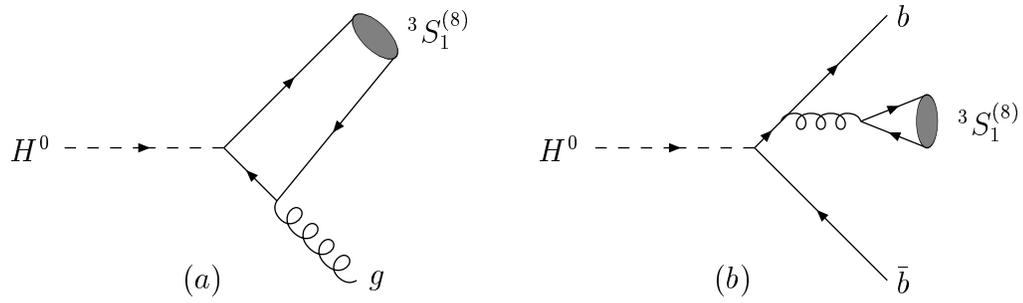
Fig.2. Color-octet processes of $H^0$ decays into quarkonia, (a) the
lowest order QCD process (b) gluon fragmentation process.\\
\vskip 0.2cm
\noindent
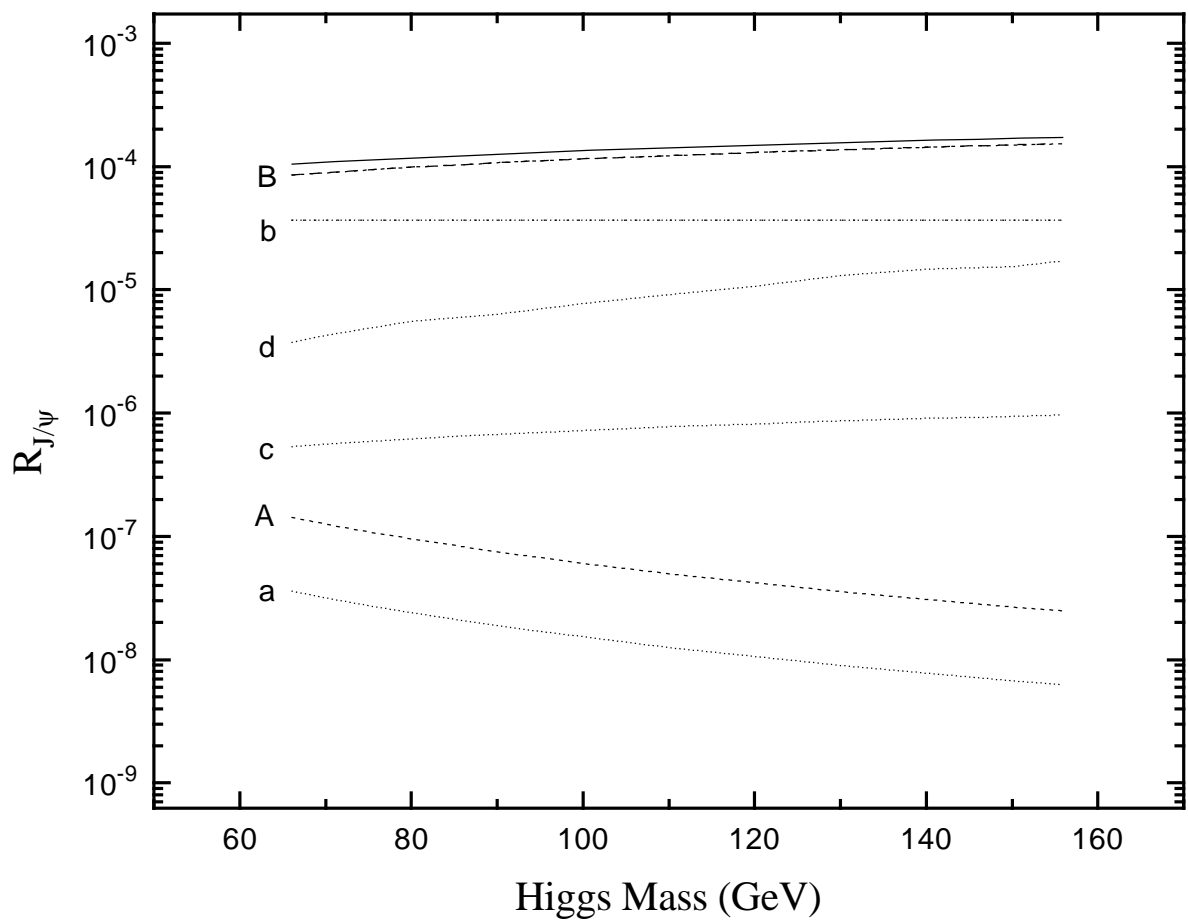
Fig.3. The total fraction ratios, relative weights of the different
$J/\psi$ production mechanisms, and different subprocesses in $H^0$
decays. The dotted curve and the dashed curve illustrate the total 
fraction ratios of the color-singlet and color-octet mechanisms,
respectively, the solid curve is a sum of them, and the subprocesses
contributions are labeled a-d and A-B corresponding to the ones in
Fig.1 and Fig.2.\\
\vskip 0.2cm
\noindent
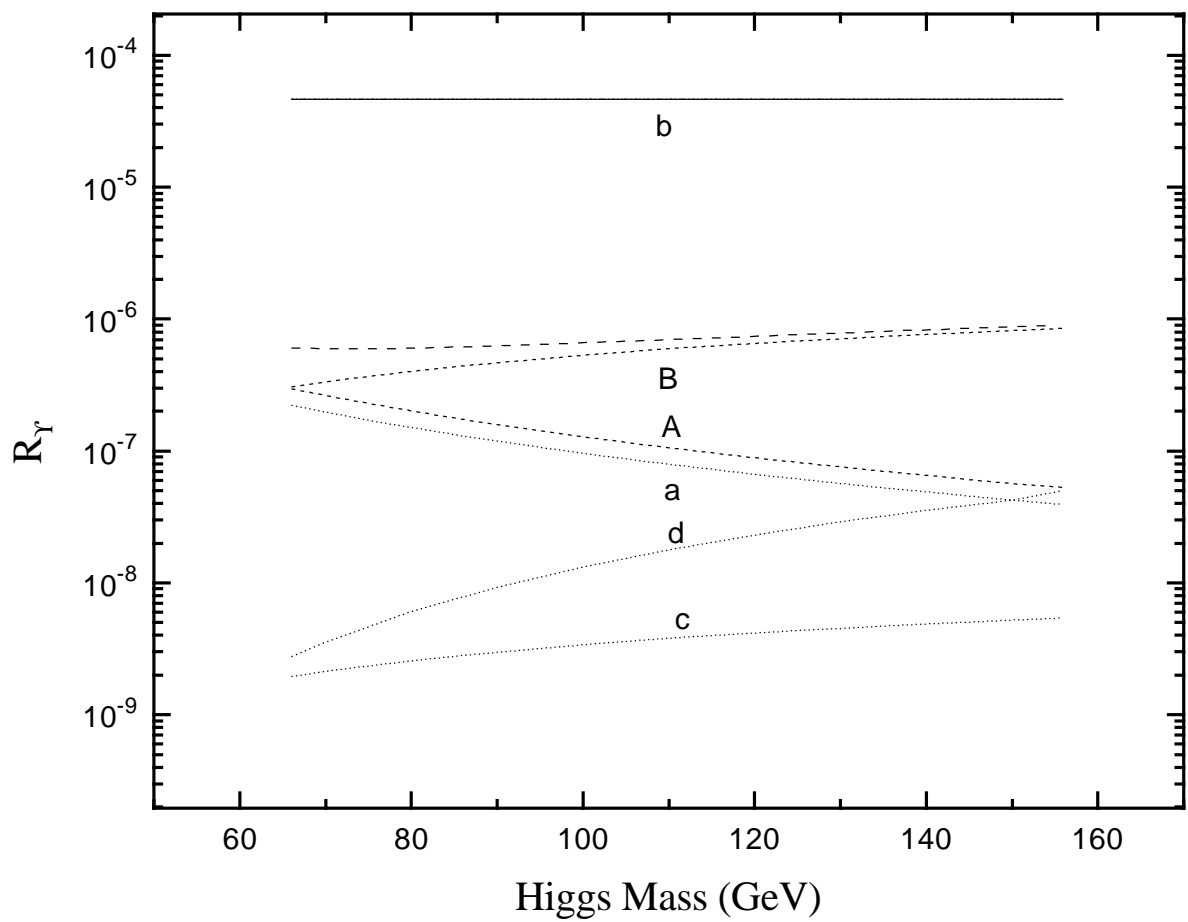
Fig.4. The total fraction ratios, relative weights of the different
$\Upsilon$ production mechanisms, and different subprocesses in $H^0$
decays. The curves in this figure are labeled the same as those in Fig.3.
\end{document}